\documentclass[twocolumn,aps,prl,showpacs]{revtex4-1}

\usepackage{amssymb}
\usepackage{mathrsfs}
\usepackage{lipsum}
\usepackage{graphicx}
\usepackage[caption=false]{subfig}
\usepackage{mathtools}
\usepackage{epstopdf}
\usepackage{xcolor}
\usepackage[colorlinks, linkcolor=red, anchorcolor=blue, citecolor=green, urlcolor=blue]{hyperref}
\DeclareGraphicsExtensions{.pdf,.jpg,.png,.eps}
\usepackage[toc,page,title,titletoc,header]{appendix}

\begin{document}
	
\title{Classical Inputs and Measurements Enable Phase Sensitivity beyond the Shot-Noise Limit}

\author{Jian-Dong Zhang}
\affiliation{School of Physics, Harbin Institute of Technology, Harbin, 150001, China}
\author{Zi-Jing Zhang}
\affiliation{School of Physics, Harbin Institute of Technology, Harbin, 150001, China}
\author{Long-Zhu Cen}
\affiliation{School of Physics, Harbin Institute of Technology, Harbin, 150001, China}
\author{Jun-Yan Hu}
\affiliation{School of Physics, Harbin Institute of Technology, Harbin, 150001, China}
\author{Yuan Zhao}
\affiliation{School of Physics, Harbin Institute of Technology, Harbin, 150001, China}

\date{\today}
	
\begin{abstract}
Coherent-state-based phase estimation is a fruitful testbed for the field of precision measurements since coherent states are robust to decoherence when compared with exotic quantum states.
The seminal work done by Caves [\href{https://doi.org/10.1103/PhysRevD.23.1693}{Phys. Rev. D \textbf{23}, 1693 (1981)}] stated that the phase sensitivity of a U(2) interferometer fed with a coherent state is limited by the shot-noise limit (SNL).
In this Letter, we demonstrate that this bound is not conclusive sensitivity limit and can be broken when the measurement includes an external phase reference.
The SNL can be surpassed by a factor of $\sqrt{2}$ and the validity is supported through the calculation of quantum Fisher information.
Additionally, we discuss other single-mode Gaussian inputs of which sensitivities are beyond the SNL.
Our work shows potential applications for many metological scenarios, particularly when the measured samples immersed in great lossy environments or can withstand bright illumination.
\end{abstract}

\maketitle

\emph{Introduction.---} Metrology is an ancient as well as modern subject committed to achieving sensitive parameter estimation.
One of the useful tools is a U(2) interferometer \cite{U2,PhysRevA.91.032103}, which can be used to estimate many slight variations on physical quantities, such as phase shifts \cite{PhysRevLett.96.010401,Giovannetti2011,PhysRevLett.104.103602,PhysRevLett.107.083601,PhysRevLett.111.033603}, polarized rotations \cite{PhysRevLett.112.103604,Schafermeier:18,PhysRevA.96.053846,Zhang:17}, and angular displacements \cite{PhysRevLett.112.200401,D2013Photonic}.
In this regard, there exists a famous theorem on sensitivity limit referred as to the SNL \cite{PhysRevD.23.1693}: \emph{The optimal sensitivity of a U(2) interferometer fed with an arbitrary single-mode state scales as $1/\sqrt{\cal N}$ with $\cal N$ being photon number inside the interferometer}.
This conclusion makes the studies on protocols based upon single-mode inputs stagnant, since all attempts are unavailing as long as one of two inputs is vacuum.

In general, we are used to proving this theorem through the use of the quantum Fisher information (QFI) \cite{PhysRevLett.72.3439}; namely, one can get ${\cal F}={\cal N}$ irrespective of photon distributions of the inputs. 
However, this result is obtained in terms of QFI calculation using a distributed anti-symmetrically operator ${\hat U_{\rm{1}}}{\rm{ = }}\exp [i{\varphi }({\hat a^\dag }\hat a - {\hat b^\dag }\hat b)/2]$, where ${{{\hat a}^\dag }}$ (${{{\hat b}^\dag }}$) and ${{{\hat a} }}$ (${{{\hat b} }}$) are the creation and annihilation operators of the path $A$ ($B$), respectively, and $\varphi$ is the estimated parameter.
In contrast to above result, if another operator ${\hat U_{\rm{2}}} = \exp (i\varphi {\hat a^\dag }\hat a)$ is used in QFI calculation, one may give ${\cal F}> {\cal N}$ regarding some inputs, indicating a sub-shot-noise-limited sensitivity.
These two operators are both physically real and introduce a relative difference in parameter  $\varphi$ between the two paths; however, they provide two different sensitivity limits.
Confusion exists as to whether the sensitivity limit given by operator $\hat U_2$ is available via a practical positive operator valued measure (POVM).

Over the past years, this seemingly counterintuitive phenomenon has got some attentions.
With inputting a coherent state combined with a squeezed vacuum, Jarzyna and Demkowicz-Dobrza\'nski did the QFI analysis and showed that two QFIs are given by different operator configurations regarding two-mode inputs \cite{PhysRevA.85.011801}.
More recently, Takeoka \emph{et al.} \cite{PhysRevA.96.052118} and You \emph{et al.} \cite{PhysRevA.99.042122} stated that the QFI calculated from $\hat U_2$ is physically achievable but one needs to deploy a POVM including phase or power reference sources. 
In particular, Takeoka \emph{et al}. demonstrated a real example to break through the SNL via a squeezed vacuum and a power reference source \cite{PhysRevA.96.052118}.

In this Letter, the potential of classical inputs and measurements to surpass the SNL is discussed.
Related to this, we report on a protocol for phase estimation with a coherent state as input.
With a POVM carrying a phase reference taken, we can obtain a sub-shot-noise-limited sensitivity suggesting an amplification effect without post-selection.
Within the reach of Gaussian inputs, we demonstrate that this effect hold true for states generated from displacement operator.
Our results have an important implication for realistic scenario where the measured samples can withstand bright illustration.

\emph{Estimation protocol.---} Consider a U(2) interferometer with an unknown phase $\varphi$, as illustrated in Fig. \ref{f1}.
A coherent state $\left| \alpha \right\rangle $ is injected into port $A$, and port $B$ is vacuum.
Accordingly, the input can be written as $\left| \psi_1  \right\rangle  = {\left| \alpha  \right\rangle _A}{\left| 0 \right\rangle _B}$.
Then the coherent state is incident on the first 50:50 beam splitter (BS), and evolves into two coherent states, $\left| \psi _2 \right\rangle  = {\left| {\alpha /\sqrt{2}} \right\rangle _A}{\left| {\alpha /\sqrt{2}} \right\rangle _B}$, where the phase introduced by the reflection of BS1 is ignored.
Subsequently, the state in path $A$ experiences a phase shift $\varphi$, the parameter we would like to estimate.
To such a unitary process there corresponds a phase operator ${\hat U_\varphi } = \exp (i\varphi {\hat a^\dag }\hat a)$. 
The phase shift is equally imprinted on each photon in path $A$, and the state becomes  $\left| \psi_3  \right\rangle  = {\left| {\alpha {e^{i\varphi }}/\sqrt{2}} \right\rangle _A}{\left| {\alpha /\sqrt{2}} \right\rangle _B}$.

From the perspective of metrology, all the devices after the phase shift are regarded as measurement.
In our protocol, the measurement can be divided into two parts. 
The first part is a beam combiner, BS2, taking the role of recombining two coherent states.
Here we set the transmissivity of BS2 to $T$ instead of a fixed value, 1/2, which is used in most of previous protocols.
As a consequence, upon leaving BS2, the reduced output of path $A$ turns to be
\begin{equation}
{\left| \psi  \right\rangle }_A = \left| {\frac{\alpha }{{\sqrt 2 }}\left( {\sqrt T {e^{i\varphi }} + \sqrt {1 - T} } \right)} \right\rangle. 
\label{e1}
\end{equation}
The second part is a balanced homodyne detection module, the output in Eq. (\ref{e1}) is superimposed on BS3 with a local oscillator $\left| \beta \right\rangle $.
Two detectors record two output intensities, and the difference between these two intensities is related to phase $\varphi$.

\begin{figure}[htbp]
	\centering\includegraphics[width=8.5cm]{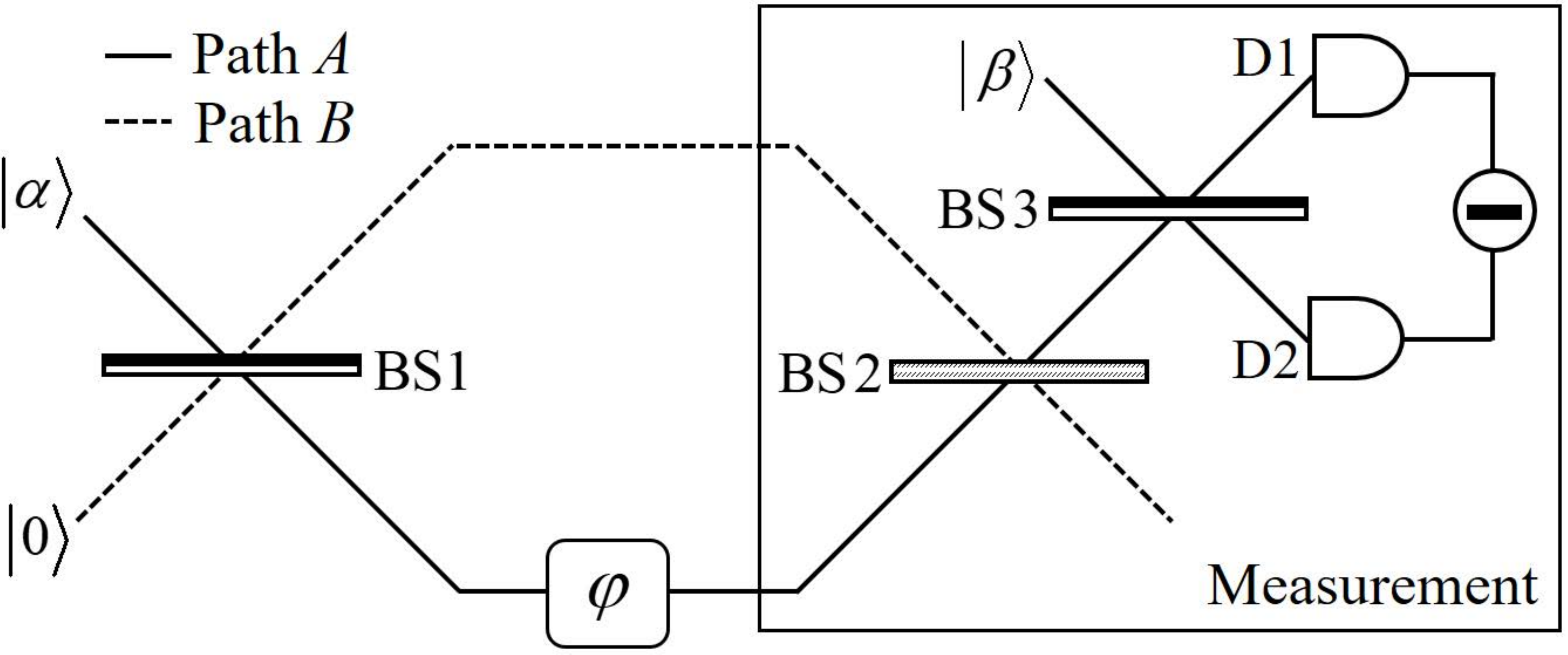}
	\caption{Schematic diagram of a U(2) interferometer composed of two BSs and an unknown phase $\varphi$.
	Here the input is a coherent state, and balanced homodyne detection is taken on the output.
	The transmissivities of BS1, BS2, and BS3 are 1/2, $T$, and 1/2, respectively.
	$\left| \beta \right\rangle $ is a strong local oscillator, i.e., a high-intensity coherent state.}
	\label{f1}
\end{figure}

\emph{Optimal sensitivity and QFI.---} We start off with the optimal sensitivity of our protocol.
With respect to the output port $A$, the operator of balanced homodyne detection is given by
${\hat X_A} = {\hat a^\dag } + \hat a$.
Combining Eq. (\ref{e1}) and measurement operator, the expectation value of the output can be calculated as:
\begin{equation}
\langle {{{\hat X}_A}}\rangle  =  - \sqrt {2T} \left| \alpha  \right|\sin \varphi, 
\label{e3}
\end{equation}
where $ \alpha = i \left| \alpha \right| $ is used throughout this work.
Further, we can give the expectation value of square of the operator,
\begin{equation}
\langle {\hat X_A^2} \rangle  = T{\left| \alpha  \right|^2}\left[ {1 - \cos \left( {2\varphi } \right)} \right] + 1.
\label{e4}
\end{equation}

Using the error propagation, phase sensitivity is found to be
\begin{align}
\Delta \varphi  = \frac{{\sqrt {\langle {\hat X_A^2} \rangle  - {\langle {\hat X_A^{}} \rangle ^2}} }}{{| {{{\partial \langle {{{\hat X}_A}} \rangle } / 
{\partial \varphi }}}|}} = \frac{1}{{| {\sqrt {2T} \alpha \cos \varphi } |}}.
\label{e5}
\end{align}
In general, an exact expression of phase sensitivity should be derived from classical Fisher information (CFI), $\Delta \varphi = 1/\sqrt{\cal{F}_{\rm c}}$.
For our protocol, we can prove the equivalence of CFI method and error propagation method, see details in Supplementary Material.
One can find that, for a fixed $T$ value, phase sensitivity, $1/{\sqrt {2T}}{{\left|  \alpha  \right|}}$, is obtained when $\varphi=0$.
This implies that the SNL can be surpassed with any $T > 1/2$.
In particular, the optimal sensitivity of our protocol after optimizing $T$ is $1/{\sqrt {2}}{{\left|  \alpha  \right|}}$, corresponding to $T=1$.

Now we turn our attention to QFI calculation.
For a two-mode separable pure state, $\rho  = {\rho _A} \otimes {\rho _B}$, the QFI can be expressed as:
\begin{equation}
{\cal F} = 4[ {\langle {{{\hat O}^2}}\rangle  - {\langle {\hat O}\rangle ^2}}],
\end{equation}
where the expectation values are taken over the state $\left| \psi_2  \right\rangle$, and the estimator $\hat O$ can be deduced from the following differential equation
\begin{equation}
\frac{{\partial {\rho _\varphi }}}{{\partial \varphi }} = i[ {\hat O,{\rho _\varphi }}]
\end{equation}
with density matrix ${\rho _\varphi } = \hat U_\varphi ^{}\left| {{\psi _2}} \right\rangle \left\langle {{\psi _2}} \right|\hat U_\varphi ^\dag$.

In our protocol, we have $\hat O = {\hat a^\dag }\hat a$; further, the QFI is calculated as:
\begin{equation}
{\cal F} = 4[ {\langle {{( {{{\hat a}^\dag }\hat a})^2}} \rangle  - {\langle {{{\hat a}^\dag }\hat a}\rangle ^2}} ] = 2{\left| \alpha  \right|^2}. 
\end{equation}

This equation means that the optimal sensitivity allowed by the QFI surpasses the SNL by a factor of $\sqrt 2$.
Meanwhile, balanced homodyne detection is the optimal strategy of our protocol as the corresponding CFI equals the QFI.
What we need to emphasize is that the QFI calculated by $\hat U_1$ holds true for POVMs without power and phase references, while that calculated by $\hat U_2$ is applicable for more general POVMs.
In addition, there exists the similar effect within the field of nonlinear phase estimation \cite{PhysRevA.72.045801,zhang2019nearly,nonlinear}.

Notice that the optimal sensitivity in our discussion is achievable with $\varphi=0$ and $T=1$.
There is no interference between the two paths, in that the condition $T=1$ amounts to removing BS2.
This is a counterintuitive and even preposterous conclusion at the first glance, since it is well known that only the relative phase, rather than phase itself, can be estimated. 
Upon further inspection, it can be seen that the state of path $A$ can interfere with the local oscillator $\left| \beta \right\rangle $.
That is, in our protocol, an external phase reference is used to provide a possibility of surpassing the SNL \cite{PhysRevA.85.011801}. 

In particular, if $\left|\beta\right|  = \left|\alpha \right|/\sqrt 2 $, our protocol shown in Fig. \ref{f1} is equivalent to a conventional balanced U(2) interferometer with intensity-difference detection.
However, it is known that phase sensitivity of a U(2) interferometer with intensity-difference detection is limited by the SNL.
Naturally, one may arise a confusing question: Why do two identical physical configurations provide two different sensitivities?
To answer this question, we make an attempt to analyze the difference between intensity-difference detection and balanced homodyne detection.

In general, a balanced homodyne detection module includes a local oscillator $\left| \beta \right\rangle $, a 50:50 BS, two detectors, and a intensity-difference processing device.
Let us consider angular momentum operators in the Schwinger representation, ${{\hat J}_x} = ( {\hat a{{\hat b}^\dag } + {{\hat a}^\dag }\hat b})/2$,
${{\hat J}_y} = {i}( {\hat a{{\hat b}^\dag } - {{\hat a}^\dag }\hat b})/2$, and	${{\hat J}_z} = ( {{{\hat a}^\dag }\hat a - {{\hat b}^\dag }\hat b} )/2$ \cite{PhysRevA.33.4033}.
They obey the SU(2) Lie algebra:
$[ {{{\hat J}_m},{{\hat J}_n}}] = i{\varepsilon _{mnk}}{\hat J_k}$,
where $\{ m,n,k\}  \in \{ x,y,z\} $, and ${\varepsilon _{mnk}}$ is the Levi-Civita symbol.

Based on these operators, the actions of BS3 and intensity-difference detection can be written as ${{\hat U}_{\rm BS}} =\exp (i\pi {{\hat J}_x}/2)$ and ${{\hat U}_{\rm D}} = 2{{\hat J}_z}$, respectively.
Using the Baker-Hausdorff lemma, we have
$\hat U_{\rm BS}^{\dag}{{\hat U}_{\rm D}^{}} \hat U_{\rm BS}^{}  = -2{\hat J_y}$.
Further, one can give the expectation value of balanced homodyne detection
\begin{equation}
\left\langle {{\psi },\beta } \right|\hat U_{\rm BS}^\dag \hat U_{\rm D}^{}\hat U_{\rm BS}^{}\left| {{\psi },\beta } \right\rangle  = \left| \beta  \right|\langle {{{\hat X}_A}} \rangle 
\end{equation}
and that of its square
\begin{equation}
\left\langle {{\psi },\beta } \right|\hat U_{\rm BS}^\dag \hat U_{\rm D}^2\hat U_{\rm BS}^{}\left| {{\psi },\beta } \right\rangle  = {\left| \beta  \right|^2}\langle {\hat X_A^2} \rangle  + \left\langle {{{\hat a}^\dag }\hat a} \right\rangle,
\end{equation}
where $\left\langle {{{\hat a}^\dag }\hat a} \right\rangle$ denotes photon number of output at port $A$.
Thus, phase sensitivity of balanced homodyne detection is expressed as: 
\begin{equation}
\Delta \varphi ' = \frac{{\sqrt {\langle {\hat X_A^2} \rangle  - {\langle {\hat X_A^{}} \rangle ^2} + \xi } }}{{| {{\partial \langle {{{\hat X}_A}} \rangle }/ {\partial \varphi }}|}}
\label{e16}
\end{equation}
with coefficient $\xi  = \langle {{{\hat a}^\dag }\hat a} \rangle /{\left| \beta  \right|^2}$.

Compared with Eq. (\ref{e5}), the numerator of Eq. (\ref{e16}) includes an extra term $\xi$, which equals the intensity ratio of the output to the local oscillator.
When $\left|\beta\right|  = \left|\alpha \right|/\sqrt 2 $, we have $\xi=1$, and then, the minimum of Eq. (\ref{e16}) is the SNL, $1/\left| \alpha  \right|$.
As increasing the intensity of the local oscillator, the coefficient $\xi$ keeps decreasing.
With a strong local oscillator taken, Eqs. (\ref{e5}) and (\ref{e16}) are approximately equal since $\langle {{{\hat a}^\dag }\hat a} \rangle$ is negligible compared to ${\left| \beta  \right|^2}$ ($\xi \simeq 0$).
This suggests that a strong local oscillator is a necessary condition regarding balanced homodyne detection, or rather, balanced homodyne detection is an approximate version derived from intensity-difference detection with strong local oscillator.

\begin{figure}[htbp]
\centering\includegraphics[width=7cm]{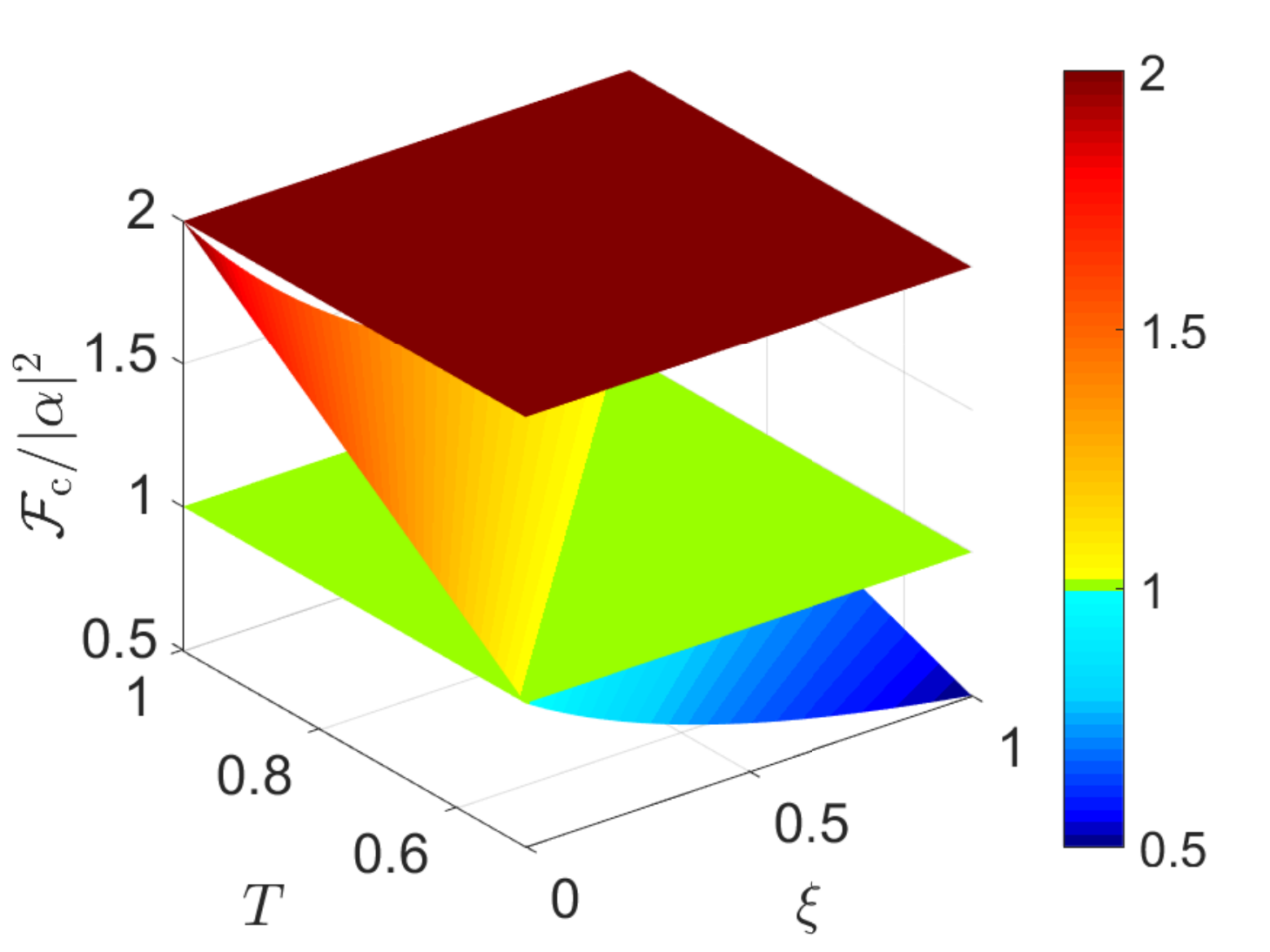}
\caption{The CFI, ${\cal F}_{\rm c}$, versus transmissivity, $T$, and coefficient, $\xi$. The upper and lower planes represent the QFI calculated by ${\hat U}_1$ ($2|\alpha|^2$) and SNL ($|\alpha|^2$). Here we implement a linear transformation, $1/|\alpha|^2$, to $z$ axis; as a result, the QFI, CFI, and SNL turn into 2, ${\cal F}_{\rm c}/|\alpha|^2$, and 1, respectively.}	
\label{f2}
\end{figure}

In Fig. \ref{f2}, we show the dependence of CFI, ${\cal F}_{\rm c}$, on transmissivity, $T$, and coefficient, $\xi$.
A distinct trend is that CFI can be improved with either the increase of transmissivity or the decrease of intensity of the local oscillator.
Meanwhile, one can find that the CFI sits at the SNL with 50:50 BS2 ($T=0.5$) and strong local oscillator ($\xi=0$), which is the result of previous studies.
Regarding a strong local oscillator, the QFI will be saturated with the CFI when transmissivity approaches 1, as shown in our protocol.

\emph{Possibility of super-sensitivity using other Gaussian states.---}
In terms of above analysis, super-sensitivity with a coherent state input is demonstrated.
Now we further discuss the potential of other Gaussian inputs to surpass the SNL.
In contrast to non-Gaussian states, Gaussian states are efficient in preparation and more robust for photon loss, making them more suitable for practical phase estimation.
Generally, a single-mode Gaussian state can always be represented as a displaced squeezed thermal state \cite{RevModPhys.84.621}.
On the basis of this fact, here we consider five kinds of states except for aforementioned coherent states: two kinds of single-parameter states and three kinds of two-parameter states (see Supplementary Material for details).

For balanced homodyne detection, the expectation values of two kinds of single-parameter states (thermal states and squeezed vacuum states) are zero, indicating that they cannot perform phase estimation. 
As to three kinds of two-parameter states---squeezed thermal states, displaced thermal states, and displaced squeezed states---the expectation value of the first kind is zero while those of the last two kinds are phase-sensitive.

The optimal sensitivity of a displaced thermal state is found to be
\begin{equation}
\Delta {\varphi _{\rm DT}} = \frac{{\sqrt {{{\cal N}_{\rm T}} + 1} }}{{\sqrt 2\left| { \alpha} \right|}},
\label{}
\end{equation}
and that of a displaced squeezed state is given by
\begin{equation}
\Delta {\varphi _{\rm DS}} = \frac{{\sqrt {{{\cosh }^2}r - \sinh r\cosh r} }}{{\sqrt 2\left| { \alpha } \right|}},
\label{e12}
\end{equation}
where ${{\cal N}_{\rm T}}$ is photon number of the initial thermal state and $r={\rm arsinh}\sqrt{{\cal N}_{\rm SV}}$ is squeezing factor with ${\cal N}_{\rm SV}$ being  photon number of the initial squeezed vacuum.

\begin{figure}[htbp]
	\centering\includegraphics[width=7.5cm]{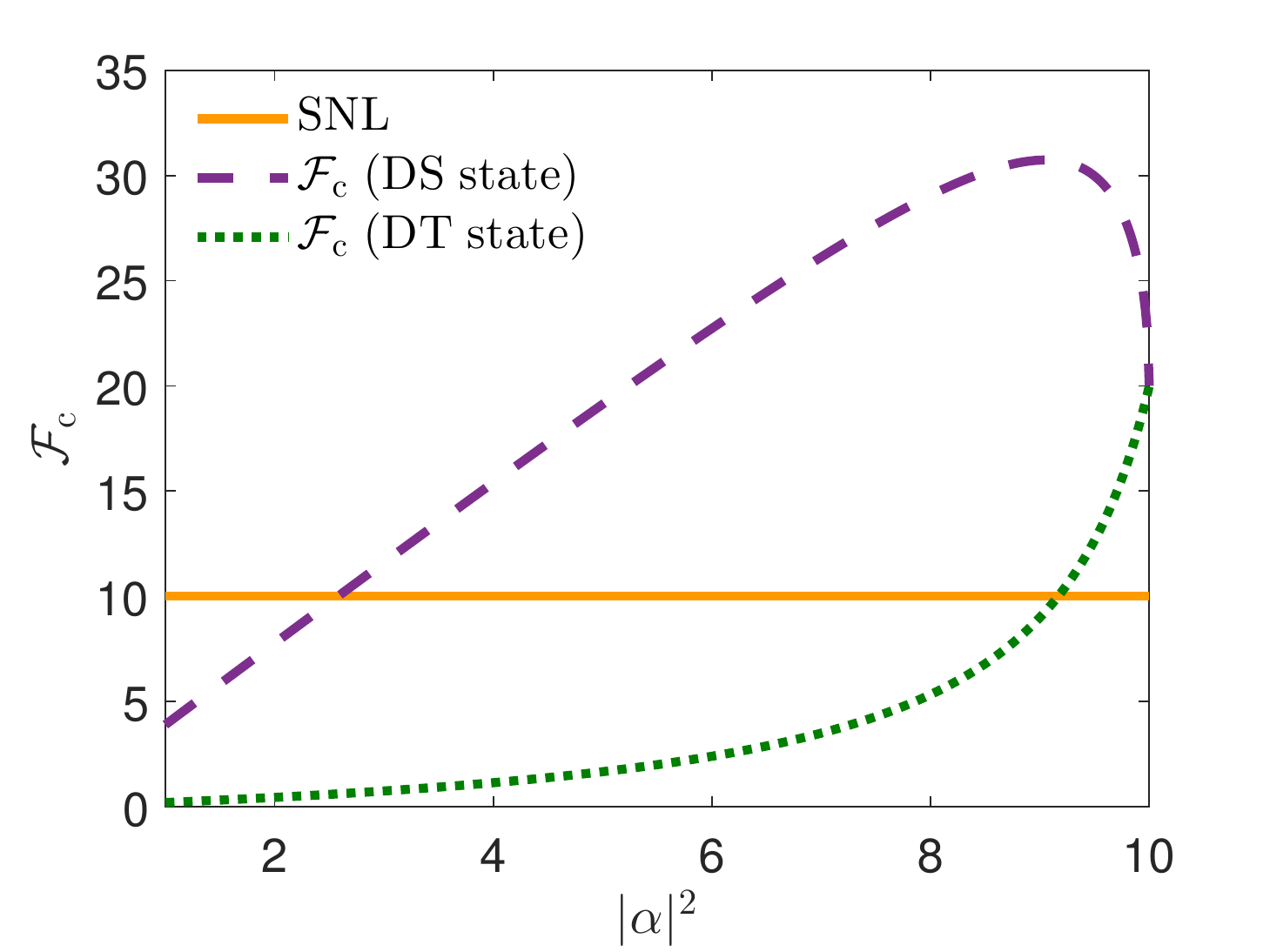}
	\caption{The CFI versus photon number originating from the operation of displacement operator, where ${\cal N}_{\rm DS}=|\alpha|^2+{\cal N}_{\rm SV}=10$ and ${\cal N}_{\rm DT}=|\alpha|^2+{\cal N}_{\rm T}=10$. ${\cal N}_{\rm SV}$ and ${\cal N}_{\rm T}$ denote the initial photon number of squeezed vacuum and that of thermal state. DS state, displaced squeezed state; DT state, displaced thermal state.}
	\label{f3}
\end{figure}

With a displaced thermal state or a displaced squeezed state as input, Fig. \ref{f3} gives CFI as a function of photon number, $|\alpha|^2$, originating from the operation of displacement operator. 
As seen in the figure, these two states can achieve sub-SNL for special proportion of $|\alpha|^2$ to total photon number.
The CFI of the displaced thermal state is a monotonically increasing function, meaning that reducing photon number of the initial thermal state increases the CFI when the total photon number is fixed.
In particular, the coherent state mentioned above is a special scenario that photon number of the initial thermal state is zero; accordingly, phase sensitivity of a displaced thermal state is inferior to that of a coherent state.

In contrast, the maximal CFI of the displaced squeezed state outperforms that of the coherent state.
That is, for our protocol, the displaced squeezed state is the optimal candidate when compared with other Gaussian inputs. 
To verify the authenticity of this advantage, we calculate QFI of the displaced squeezed state,
\begin{equation}
{\cal F} =( {{e^{2r}} + 1} ){\left| \alpha  \right|^2} + \frac{1}{2}{\sinh ^2}\left( 2r\right)  + {\sinh ^2}r.
\label{e18}
\end{equation}
Equation (\ref {e18}) indicates the CFI beyond the SNL is achievable, since ${\cal F}\geqslant {{\cal F}_{\rm c}}= 1/\Delta \varphi^2_{\rm DS}$ for any values of $|\alpha|^2$.

It can be seen that the CFI of the displaced squeezed state is a non-monotonic concave function.
This means that there exists an optimal photon number for the initial squeezed vacuum to  maximize the CFI regarding a fixed total photon number. 
In order to obtain the maximal CFI, we differentiate the CFI with respect to $|\alpha|^2$ and then let the result equal to zero.
The corresponding solution is given by
\begin{equation}
{\left| \alpha  \right|^2} = \frac{{2( {1 + 3{\cal N}_{\rm DS}^{} + 2{\cal N}_{\rm DS}^2 - \sqrt {1 + 3{\cal N}_{\rm DS}^{} + 3{\cal N}_{\rm DS}^2 + {\cal N}_{\rm DS}^3} })}}{{3{\rm{ + }}4{{\cal N}_{\rm DS}}}}.
\label{e14}
\end{equation}
Consequently, for a displaced squeezed state with ${\cal N}_{\rm DS}$ photons on average, ${\cal N}_{\rm DS}-|\alpha|^2$ is the optimal photon number of the initial squeezed vacuum achieving the maximal CFI.
To a large ${\cal N}_{\rm DS}$ there corresponds a fact that $|\alpha|^2 \simeq {\cal N}_{\rm DS}$ and $r \simeq 0$.
Such a requirement for squeezing factor indicates that it may be workable to input a displaced squeezed state carrying large photon number with current technology.
Finally, we consider the extent of maximal CFI superior to SNL.
By substituting Eq. (\ref{e14}) into Eq. (\ref{e12}), we have the maximal CFI
\begin{equation}
\max \left[ {{{\cal F}_{\rm{c}}}} \right] = \frac{{4 + 12{\cal N}_{\rm DS}^{} + 8{\cal N}_{\rm DS}^2 - 4\sqrt {{{\left( {1 + {\cal N}_{\rm DS}^{}} \right)}^3}} }}{{1 + {\cal N}_{\rm DS}^{} + 2\sqrt {{{\left( {1 + {\cal N}_{\rm DS}^{}} \right)}^3}}  - \sqrt {{\cal H}_1{\cal H}_2} }}
\label{}
\end{equation}
with
\begin{eqnarray}
{\cal H}_1= &&2\sqrt {{{\left( {1 + {\cal N}_{\rm DS}^{}} \right)}^3}}  - 3{\cal N}_{\rm DS}^{} - 2,\\
{\cal H}_2= &&2\sqrt {{{\left( {1 + {\cal N}_{\rm DS}^{}} \right)}^3}}  + {\cal N}_{\rm DS}^{} + 1.
\end{eqnarray}

\begin{figure}[htbp]
	\centering\includegraphics[width=7.5cm]{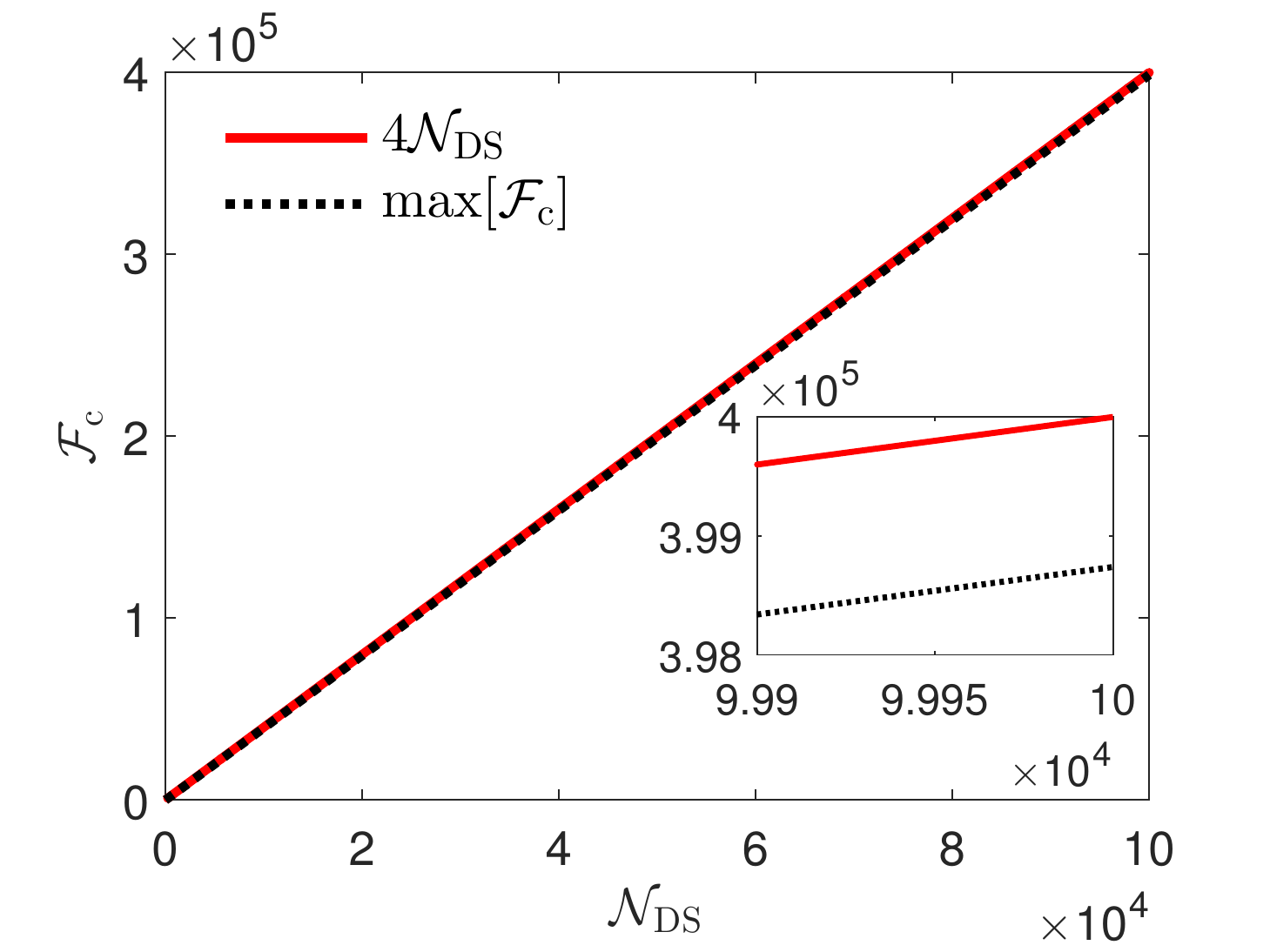}
	\caption{The maximal CFI versus photon number of displaced squeezed state, where the solid line, 4${\cal N}_{\rm DS}$, is a reference, and the inset shows a zoom of the graph within high-photon interval.}
	\label{f4}
\end{figure}

Figure \ref{f4} shows the maximal CFI against photon number, ${\cal N}_{\rm DS}$.
There is a nearly linear relationship between them with a factor of $\sim$4.
We plot the reference line of $4{\cal N}_{\rm DS}$ for comparison, and find that they approximately overlap with each other.
These results reveal that phase sensitivity of a displaced squeezed state is roughly twice of SNL in high-photon region.

\emph{Conclusion.---} In summary, we have proposed a protocol used for phase estimation, with a coherent state as input and balanced homodyne detection as measurement.
The optimal sensitivity achieves the limit predicted by QFI, and can surpass the SNL by a factor of $\sqrt 2$ in the absence of probabilistic pre- and post-selections. 
We discuss other single-mode Gaussian states, and reveal that our protocol is applicable to those states arising from displacement operation.
In particular, displaced squeezed states emerge as the optimal candidate in that they outperform other Gaussian states and surpass the SNL by a factor of $\sim$2. 
We believe that our work paves a way for the realization of super-sensitive phase estimation by single-mode Gaussian state engineering.

\emph{Acknowledgment.---} This work was supported by National Natural Science Foundation of China (Grant No. 61701139).

\begin{widetext}

\section*{Supplementary Material}

\section{A: Optimal sensitivity calculation based on the classical Fisher information}

In this section we give the optimal sensitivity of our protocol with the classical Fisher information (CFI) calculation.
In general, it is complicated to calculate an analytic expression of CFI regarding intensity-based detection.
Here we provide two methods for the CFI calculation.
The first method is only suitable for calculating maximum CFI, but it has no limitation on the output state.
By contrast, the second one can show variation of the CFI with respect to the estimated parameter, while it merely holds true for the state satisfying particular probability distribution.   
In what follows, we direct our attention to the first method and focus on phase $\varphi$ in the vicinity of the origin \cite{PhysRevA.72.045801}.

Since $X$-quadrature is detected in our protocol, for a given phase $\varphi$, the probability of outcome $x$ is found to be 
\begin{equation}
P\left( {x\left| {\kern 1pt}\varphi  \right.} \right) = {\left| {\left\langle {x}
		{\left | {\vphantom {x {{\psi _A}}}}
			\right. \kern-\nulldelimiterspace}
		{{{\psi _A}}} \right\rangle_{T=1} } \right|^2} = {\left| {\left\langle x \right|\exp \left( {i\varphi {{\hat a}^\dag }\hat a} \right)\left| {\frac{\alpha }{{\sqrt 2 }}} \right\rangle } \right|^2}, 
\end{equation}
where $\left| x \right\rangle$ is the eigenvector of the operator $\hat x$ with eigenvalue $x$.

Further, the CFI can be expressed as: 
\begin{equation}
{{\cal F}_{\rm{c}}} = {\int_{ - \infty }^\infty {\frac{1}{{P\left( {x\left| {\kern 1pt} \varphi  \right.} \right)}}\left[ {\frac{{\partial P\left( {x\left| {\kern 1pt} \varphi  \right.} \right)}}{{\partial \varphi }}} \right]} ^2 dx}.
\label{2}
\end{equation}

We consider the differential term at the phase origin
\begin{align}
\nonumber {\left. {\frac{\partial }{{\partial \varphi }}\left\langle x \right|\exp \left( {i\varphi {{\hat a}^\dag }\hat a} \right)\left| {\frac{\alpha }{{\sqrt 2 }}} \right\rangle } \right|_{\varphi  = 0}} &= i\left\langle x \right|{{\hat a}^\dag }\hat a\left| {\frac{\alpha }{{\sqrt 2 }}} \right\rangle \\
\nonumber &=  - \frac{{\left| \alpha  \right|}}{2}\left( {x - \frac{\partial }{{\partial x}}} \right)\left\langle {x}
{\left | {\vphantom {x {\frac{\alpha }{{\sqrt 2 }}}}}
	\right. \kern-\nulldelimiterspace}
{{\frac{\alpha }{{\sqrt 2 }}}} \right\rangle \\
&= \frac{{\left| \alpha  \right|}}{{\sqrt 2 }}\left( {i\left| \alpha  \right| - \sqrt 2 x} \right)\left\langle {x}
{\left | {\vphantom {x {\frac{\alpha }{{\sqrt 2 }}}}}
	\right. \kern-\nulldelimiterspace}
{{\frac{\alpha }{{\sqrt 2 }}}} \right\rangle, 
\label{3}
\end{align}
where ${\hat a^\dag } = (\hat x - i\hat p)/\sqrt 2 $ and $(\hat x - i\hat p)\left| x \right\rangle  = (x - \frac{\partial }{{\partial x}})\left| x \right\rangle $ are used, and the wave function of a coherent state $\left| \alpha \right\rangle $ in position representation is given by

\begin{equation}
\left\langle {x}
{\left | {\vphantom {x \alpha }}
	\right. \kern-\nulldelimiterspace}
{\alpha } \right\rangle  = {\pi ^{ - 1/4}}\exp \left( { - \frac{{{x^2}}}{2} - \frac{{{{\left| \alpha  \right|}^2}}}{2} + \sqrt 2 x \alpha  - \frac{{{\alpha ^2}}}{2}} \right)
\label{4}
\end{equation}

Based on Eqs. (\ref{3}) and (\ref{4}), we get
\begin{align}
{\left. {\frac{{\partial P\left( {x\left| {\kern 1pt} \varphi  \right.} \right)}}{{\partial \varphi }}} \right|_{\varphi  = 0}} = {i\left\langle x \right|{{\hat a}^\dag }\hat a\left| {\frac{\alpha }{{\sqrt 2 }}} \right\rangle } \left\langle {{\frac{\alpha }{{\sqrt 2 }}}}
{\left | {\vphantom {{\frac{\alpha }{{\sqrt 2 }}} x}}
	\right. \kern-\nulldelimiterspace}
{x} \right\rangle  + {\rm H.c.}=  - 2x\left| \alpha  \right|P\left( {x\left| {\kern 1pt} 0 \right.} \right)
\label{5}
\end{align}
with 
\begin{equation}
P\left( {x\left| {\kern 1pt} 0 \right.} \right) = \frac{1}{{\sqrt \pi  }}{e^{ - {x^2}}},
\label{6}
\end{equation}
where ${\rm H.c.}$ stands for Hermitian conjugation.

Combining Eqs.(\ref{5}) and (\ref{6}), we can calculate the maximal CFI as
\begin{equation}
{\left. {{{\cal F}_{\rm{c}}}} \right|_{\varphi  = 0}} = \int_{ - \infty }^\infty  {\frac{4}{{\sqrt \pi  }}{{\left| \alpha  \right|}^2}{x^2}{e^{ - {x^2}}}dx} \\
= 2{\left| \alpha  \right|^2}
\label{7}
\end{equation}
through the use of an integral formula
\begin{equation}
\int_0^\infty  {{x^{2k}}\exp \left( { - \frac{{{x^2}}}{{{a^2}}}} \right)dx}  = \sqrt \pi  \frac{{\left( {2k} \right)!}}{{k!}}{\left( {\frac{a}{2}} \right)^{2k + 1}}.
\label{8}
\end{equation}

Equation (\ref{7}) indicates that the optimal sensitivities calculated from the CFI and error propagation are the same.

Now we turn our attention to the second method.
One can find that, in our protocol, the output is a coherent state.
Therefore, the conditional probability $P\left( {x\left| {\kern 1pt}\varphi  \right.} \right) $ follows Gaussian distribution 
\begin{equation}
P\left( {x\left| {\kern 1pt} \varphi  \right.} \right) = \frac{1}{{\sqrt \pi  }}\exp \left[ { - {{\left( {x - \sqrt 2 \left| \alpha  \right|\cos \varphi } \right)}^2}} \right],
\label{9}
\end{equation}
as illustrated in Fig. \ref{f1}. 

Based on the expression in Ref. \cite{PhysRevA.93.023810}, the CFI is found to be
\begin{equation}
{{\cal F}_{\rm{c}}} = \frac{{{{\left( {\frac{\partial }{{\partial \varphi }}\langle {{{\hat X}_A}} \rangle } \right)}^2} + 2{{\left( {\frac{\partial }{{\partial \varphi }}\Omega } \right)}^2}}}{{{\Omega ^2}}}
\label{e10}
\end{equation}
with variance of the measurement operator
\begin{equation}
{\Omega ^2} = \langle {\hat X_A^2} \rangle  - {\langle {{{\hat X}_A}}\rangle ^2}.
\end{equation}
Notice that Eq. (\ref{e10}) develops into the error propagation when the variance is not a function of the estimated phase.

\begin{figure}[htbp]
	\centering\includegraphics[width=8cm]{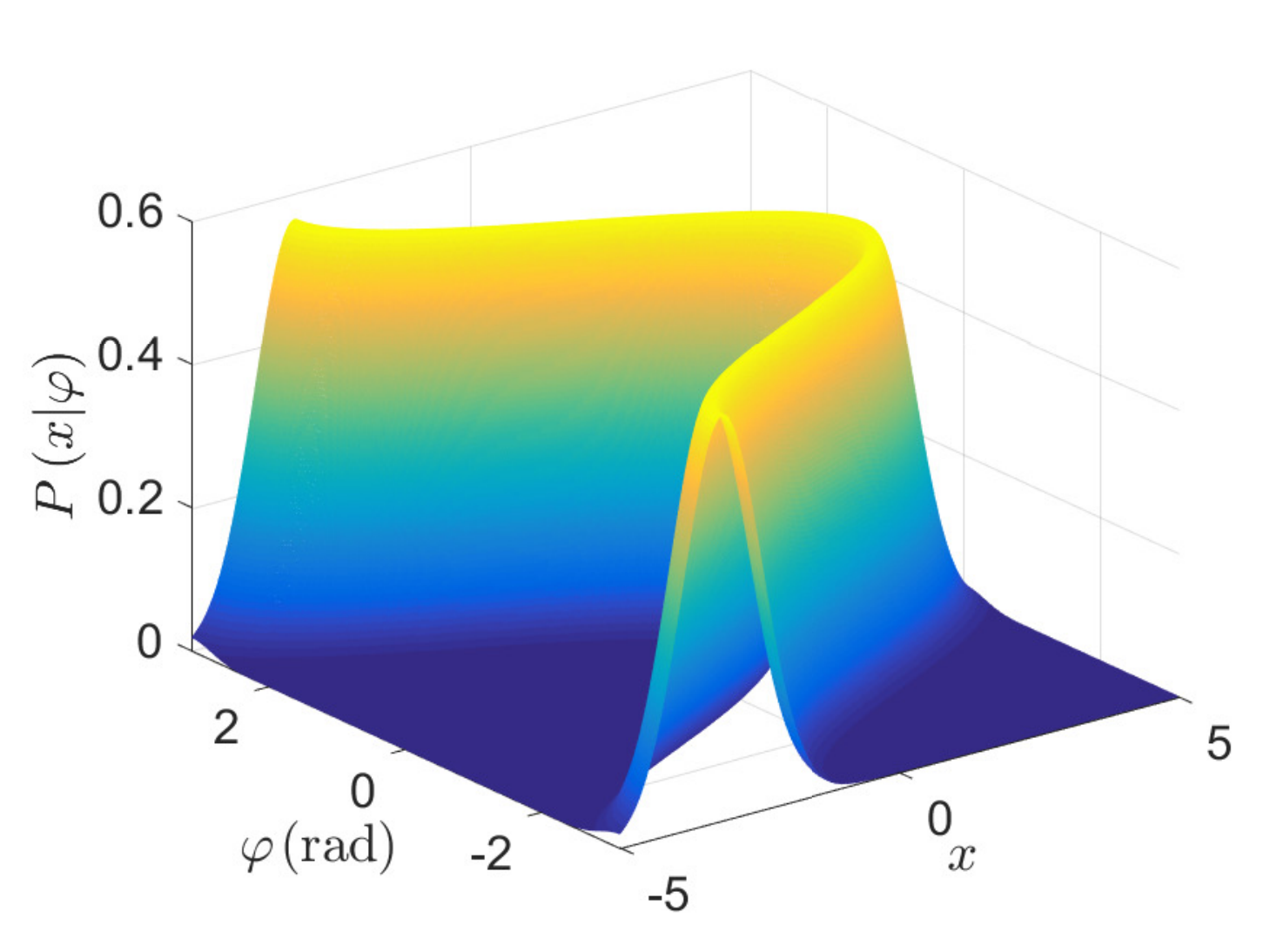}
	\caption{The probability distribution $	P\left( {x | \varphi  } \right) $ of a coherent state $\left| \alpha \right\rangle $ ($\alpha = \left| \alpha \right|e^{i\varphi}$) against outcome $x$ and phase $\varphi$, where the mean photon number $\left| \alpha \right|^2 = 10$.}
	\label{f1}
\end{figure}

\section{B: Sensitivities of other Gaussian states with balanced homodyne detection}

Here we make use of the Heisenberg picture to discuss the potential of other single-mode Gaussian states to surpass the SNL. 
We start off with the square of measurement operator, which can be represented as	
\begin{equation}
\hat X_A^2 = \hat a_2^\dag \hat a_2^\dag  + \hat a_2^{}\hat a_2^{} + 2\hat a_2^\dag \hat a_2^{} + 1
\label{e12}
\end{equation}
with $\hat a_2^\dag$ and $\hat a_2^{}$ being the creation and annihilation operators for output port $A$.

We can write transformation of BS to mode operators as follows:
\begin{equation}
\left( {\begin{array}{*{20}{c}}
	{\hat a_2^{}}\\
	{\hat b_2^{}}
	\end{array}} \right) =  \frac{1}{{\sqrt 2  }}\left( {\begin{array}{*{20}{c}}
	1&i\\
	i&1
	\end{array}} \right)\left( {\begin{array}{*{20}{c}}
	{\hat a_1^{}}\\
	{\hat b_1^{}}
	\end{array}} \right),
\label{}
\end{equation}
where $\hat a_1^\dag$ and $\hat a_1^{}$ denote the creation and annihilation operators for input port $A$.

Based on this transformation, the expectation value of each term in Eq. (\ref{e12}) turns out to be
\begin{align}
\langle {\hat a_2^\dag \hat a_2^{}} \rangle  &= \frac{1}{2}\left\langle \psi  \right|\hat a_1^\dag \hat a_1^{}\left| \psi  \right\rangle, 
\label{e14}\\
\left\langle {\hat a_2^{}\hat a_2^{}} \right\rangle  &= \frac{1}{2}{e^{i2\varphi }}\left\langle \psi  \right|\hat a_1^{}\hat a_1^{}\left| \psi  \right\rangle, \\
\langle {\hat a_2^\dag \hat a_2^\dag } \rangle &= \frac{1}{2}{e^{ - i2\varphi }}\left\langle \psi  \right|\hat a_1^\dag \hat a_1^\dag \left| \psi  \right\rangle,
\label{e16}
\end{align}
where $\left| \psi  \right\rangle$ is the input state.

Combining Eqs. (\ref{e14})-(\ref{e16}), we get the expectation value of square of measurement operator, 
\begin{equation}
\langle {\hat X_A^2} \rangle  = \frac{1}{2}\left( {{e^{i2\varphi }}\left\langle \psi  \right|\hat a_1^{}\hat a_1^{}\left| \psi  \right\rangle  + {e^{ - i2\varphi }}\left\langle \psi  \right|\hat a_1^\dag \hat a_1^\dag \left| \psi  \right\rangle } \right) + \left\langle \psi  \right|\hat a_1^\dag \hat a_1^{}\left| \psi  \right\rangle  + 1.
\label{}
\end{equation}
Similarly, the expectation value of measurement operator is given by

\begin{equation}
\langle {\hat X_A^{}} \rangle  = \frac{1}{{\sqrt 2 }}\left( {{e^{ - i\varphi }}\left\langle \psi  \right|\hat a_1^\dag \left| \psi  \right\rangle  + {e^{i\varphi }}\left\langle \psi  \right|\hat a_1^{}\left| \psi  \right\rangle } \right).
\label{}
\end{equation}

Now we move on to a general single-mode Gaussian state, given by 

\begin{equation}
{\rho _{\rm G}} = \hat D\left( \alpha  \right)\hat S\left( r \right){\rho _{\rm T}}{\hat S^\dag }\left( r \right){\hat D^\dag }\left( \alpha  \right),
\label{e21}
\end{equation}
with a thermal state
\begin{equation}
{\rho _{\rm T}} = \sum\limits_m^\infty  {\frac{{{\cal N}_{\rm T}^m}}{{{{\left( {{{\cal N}_{\rm T}} + 1} \right)}^{m + 1}}}}\left| m \right\rangle \left\langle m \right|}, 
\end{equation}		
where ${\cal N}_{\rm T}$ is the mean photon number of the thermal state, $\hat D\left( \alpha  \right)$ and $\hat S\left( r \right)$ are displacement operator and squeeze operator, respectively.

The actions of these two operators on $\hat a_1^{}$ and $\hat a_1^\dag$ are given by 
\begin{align}
{{\hat S}^\dag }\left( r \right){{\hat a}_1^{}}\hat S\left( r \right) &= {{\hat a}_1^{}}\cosh r - e^{i\theta}\hat a_1^\dag \sinh r, \\
{{\hat S}^\dag }\left( r \right)\hat a_1^\dag \hat S\left( r \right) &= \hat a_1^\dag \cosh r - e^{-i\theta}{{\hat a}_1^{}}\sinh r,
\end{align}
and
\begin{align}
{{\hat D}^\dag }\left( \alpha \right){{\hat a}_1}\hat D\left( \alpha \right) &= {{\hat a}_1} + \alpha, \\
{{\hat D}^\dag }\left( \alpha \right)\hat a_1^\dag \hat D\left( \alpha \right) &= \hat a_1^\dag  + \alpha^{*}. 
\end{align}

In terms of Eq. (\ref{e21}), there are five kinds of states are discussed in the following.

(I) Thermal states

The first kind of state we consider is thermal states ($r = 0$, $\left| \alpha \right| = 0 $). 
One can find that the expectation value of measurement operator is
\begin{equation}
{{\langle {\hat X_A^{}} \rangle} _{\rm T}} = 0.
\label{e26}
\end{equation}
Equation (\ref{e26}) suggests that our measurement strategy cannot be used for this kind of state as the expectation value is not a function of the estimated phase.

(II) Squeezed vacuum states

Next, we consider the second kind of state: squeezed vacuum ($n_{\rm T} = 0$, $\left| \alpha \right| = 0 $). 
Unfortunately, the expectation value of measurement operator is found to be
\begin{equation}
{{\langle {\hat X_A^{}}\rangle }_{\rm SV}} = 0,
\label{}
\end{equation}
which contains no information on the estimated phase.

(III) Squeezed thermal states

The third kind of state is squeezed thermal states ($\left| \alpha \right| = 0 $). 
We calculate the expectation value of measurement operator as
\begin{equation}
{{\langle {\hat X_A^{}} \rangle} _{\rm ST}} = 0.
\label{}
\end{equation}
This means that our measurement strategy is also not suitable for this kind of state.

(IV) Displaced thermal states

The fourth kind of state is displaced thermal states ($ r = 0 $). 
In contract to above three kinds of states, the expectation value of measurement operator is a phase-sensitive function, given by
\begin{equation}
{{\langle {\hat X_A^{}} \rangle} _{\rm DT}} = \frac{1}{{\sqrt 2 }}\left( {{e^{ - i\varphi }}{\alpha ^ * } + {e^{i\varphi }}\alpha } \right).
\label{}
\end{equation}
Further, we have
\begin{equation}
{{\langle {\hat X_A^2} \rangle} _{\rm DT}} = \frac{1}{2}\left[ {{e^{ - i2\varphi }}{{\left( {{\alpha ^ * }} \right)}^2} + {e^{i2\varphi }}{\alpha ^2}} \right] + {{\cal N}_{\rm DT}} + 1,
\label{}
\end{equation}
where ${{\cal N}_{\rm DT} = {\cal N}_{\rm T}} + |\alpha|^2$ is the mean photon number of the displaced thermal state with $|\alpha|^2$ being the photon number which originates from the operation of displacement operator. 
The corresponding phase sensitivity can be calculated as
\begin{equation}
\Delta {\varphi _{\rm DT}} = \frac{{\sqrt {{{\cal N}_{\rm T}} + 1} }}{{\left| {\sqrt 2 \alpha \cos \varphi } \right|}}
\label{}
\end{equation}

It is obvious that increasing $|\alpha|^2$ can improve phase sensitivity when the total photon number is fixed.
By solving the following equation,
\begin{equation}
\max \left[ {{\Delta ^2}{\varphi _{\rm DT}}} \right] = \frac{1}{{{{\cal N}_{\rm T}} + {{\left| \alpha  \right|}^2}}},
\label{}
\end{equation}
we can obtain the threshold for $|\alpha|^2$,
\begin{equation}
{\left| \alpha  \right|}^2 = \frac{{{\cal N}_{\rm T}^2 + {\cal N}_{\rm T}^{}}}{{1 - {\cal N}_{\rm T}^{}}}.
\label{e32}
\end{equation}
That is, only when the value of $|\alpha|^2$ is more than the threshold in Eq. (\ref{e32}) can phase sensitivity surpass the SNL.

(V) Displaced squeezed states

The last kind of state is displaced squeezed states ($\left| \alpha \right| = 0 $). 
We can find that the expectation value of measurement operator is dependent of the estimated phase,
\begin{equation}
{{\langle {\hat X_A^{}} \rangle} _{\rm DS}} = \frac{1}{{\sqrt 2 }}\left( {{e^{ - i\varphi }}{\alpha ^ * } + {e^{i\varphi }}\alpha } \right).
\label{e33}
\end{equation}
Further, the expectation value of square of measurement operator is given by
\begin{equation}
{\langle {\hat X_A^2} \rangle}_{\rm DS}  = \frac{1}{2}\left\{ {{e^{-i2\varphi }}\left[ {{{\left( {{\alpha ^ * }} \right)}^2} - {e^{ - i\theta }}\sinh r\cosh r} \right] + {e^{ i2\varphi }}\left( {{\alpha ^2} - {e^{ i\theta }}\sinh r\cosh r} \right)} \right\} + {{\cal N}_{\rm DS}} + 1
\label{e34}
\end{equation}
where ${{\cal N}_{\rm DS} = {\cal N}_{\rm SV}} + |\alpha|^2$ is the mean photon number of the displaced squeezed state with $ {{\cal N}_{\rm SV}} = {{\sinh }^2}r$.
Similarly, $|\alpha|^2$ stands for the photon number originating from the operation of displacement operator.
Based on Eqs. (\ref{e33}) and (\ref{e34}), we can obtain the phase sensitivity
\begin{equation}
\Delta {\varphi _{\rm DS}} = \frac{{\sqrt {{{\cosh }^2}r - \sinh r\cosh r\cos \left( {2\varphi  + \theta } \right)} }}{{\left| {\sqrt 2 \alpha \cos \varphi } \right|}}
\label{}
\end{equation}
By solving the following equation,
\begin{equation}
\max \left[ {{\Delta ^2}{\varphi _{\rm DS}}} \right] = \frac{1}{{{{\sinh }^2}r + {{\left| \alpha  \right|}^2}}},
\label{}
\end{equation}
we get the threshold for $|\alpha|^2$,
\begin{equation}
{\left| \alpha  \right|}^2 = \frac{{{{\cal N}_{\rm SV}}\left( {{{\cal N}_{\rm SV}} + 1 - \sqrt {{{\cal N}_{\rm SV}}\left( {{{\cal N}_{\rm SV}} + 1} \right)} } \right)}}{{1 - {{\cal N}_{\rm SV}} + \sqrt {{{\cal N}_{\rm SV}}\left( {{{\cal N}_{\rm SV}} + 1} \right)} }}.
\label{e38}
\end{equation}
Equation (\ref{e38}) indicates that one can achieve phase sensitivity beyond the SNL with any values of $|\alpha|^2$ in excess of this threshold.

\end{widetext}


\begin{thebibliography}{24}
\bibitem [{U2()}]{U2}%
	\BibitemOpen
	\href@noop {} {\ }\bibinfo {note} {The phase transformations for SU(2), U(1),
		and U(2) interferometers can be described as: $\exp [i({\hat a^\dag }\hat a -
		{\hat b^\dag }\hat b){\phi _ d}/2]$, a phase shift distributed
		anti-symmetrically in two paths; $\exp [i({\hat a^\dag }\hat a + {\hat b^\dag
		}\hat b){\phi _ s }]$, a phase shift equally imprinted on two paths; and
		$\exp [i({\hat a^\dag }\hat a{\phi _1} + {\hat b^\dag }\hat b{\phi _2})]$,
		two different phase shifts in two paths; respectively. With variable
		substitutions $\phi_1=({\phi _ s}+{\phi _ d})/2$ and $\phi_2=({\phi _
			s}-{\phi _ d})/2$, the group relationship is found to be $\rm {U(2)} = \rm
		{U(1)} \times \rm {SU(2)}$.}\BibitemShut {Stop}%
\end{thebibliography}
\end{document}